\documentclass[
  reprint,
  amsmath,amssymb,
  pra,
  aps
]{revtex4-2}
\usepackage{graphicx}
\usepackage{xcolor}
\usepackage{tikz}
\usetikzlibrary{patterns, decorations.pathreplacing, calc, positioning, arrows.meta, backgrounds}
\usepackage{pgfplots}
\pgfplotsset{compat=1.18}
\newcounter{panel}[figure]
\renewcommand{\thepanel}{\the\numexpr\value{figure}+1\relax(\alph{panel})}
\newcommand{\panel}[1]{\refstepcounter{panel}\label{#1}\textbf{(\alph{panel})}}

\begin{document}

\title{Real-time decoder for a MegaQuOp quantum computer using a single CPU}
\author{Min Ye, Andrii Maksymov, Nicolas Delfosse}
\affiliation{IonQ Inc.}
\date{\today}

\begin{abstract}
As quantum computers advance toward the regime of MegaQuOp machines executing millions of gates, a decoding system capable of real-time error correction in such a device will be crucial. Recent efforts have been focused on decoding an error-corrected memory or a small number of logical operations. Here we demonstrate an end-to-end real-time decoding stack for a universal fault-tolerant trapped-ion quantum computer architecture capable of decoding real workloads with millions of logical gates over hundreds of logical qubits. The complete pipeline, including on the fly detector error model generation, decoding of all logical qubits, logical operations, and magic-state factories, runs on a single CPU. We benchmark the decoder on practically relevant quantum applications spanning up to 408 logical qubits, and up to one million $T$ gates. Assuming a trapped-ion architecture with 1 to 5 ms cycle time, the decoding delay stretches the computation by less than $0.3\%$ at $p_{\mathrm{CNOT}}=10^{-4}$ and less than $12\%$ at $p_{\mathrm{CNOT}}=5\times 10^{-4}$ for all workloads studied. These results demonstrate real-time decoding at MegaQuOp scale on a single conventional CPU.
\end{abstract}

\maketitle

\section{Introduction}
The realization of a MegaQuOp quantum computer, that is a device capable of executing millions of operations on hundreds of logical qubits, represents a major milestone for the quantum computing community and would provide a new tool for scientific exploration~\cite{preskill2025beyond}.
However, real-time decoding of hundreds of logical qubits throughout a quantum computation comprising millions of logical operations remains a major challenge~\cite{battistel2023real}.

We consider the problem of decoding a universal logical instruction set including memory blocks, magic state factories, logical measurements and logical unitary gates.
To allow for magic state injection and conditional logical operations, the decoding must be performed in a streaming fashion to extract logical measurement outcomes at runtime.
Moreover, to adjust the decoder to the branches selected during conditional operations, its input data, the so-called detector error model (DEM)~\cite{gidney2021stim}, must be generated on-the-fly, which is non-trivial for quantum architectures performing logical operations by merging memory blocks with complex ancilla patches~\cite{yoder2025tour, webster2026pinnacle, cain2026shor}.
Decoding logical CNOTs is also a non-trivial task~\cite{zhou2025low, sahay2025error, turner2026scalable, gu2026color, stack2026transversal}.
Finally, the syndrome data must be processed by the decoding stack faster than it is generated to avoid a decoding backlog that can induce an exponential slowdown of the quantum computation~\cite{terhal2015quantum}. 
Although decoders have been extensively optimized for a memory block and for some logical operations, no end-to-end real-time decoding of a universal quantum computation at scale has been demonstrated.

Previous work focuses on increasing the speed and throughput of surface code and quantum low-density parity-check (LDPC) memory decoders in software~\cite{higgott2022pymatching, wu2023fusion} and using specialized hardware such as FPGAs~\cite{liyanage2023scalable, ziad2025local, valentino2025quekuf, maurer2025real, maurya2025fpga, bascones2026scalable}, TPUs~\cite{senior2025scalable}, GPUs~\cite{gu2026scalable} and ASICs~\cite{barber2025real} to meet the microsecond latency requirements of superconducting quantum computers.
Offline decoding of 10 transversal CNOTs is reported in~\cite{wan2024iterative}, reaching 64 transversal CNOTs in~\cite{cain2024correlated, cain2025fast}, 
760 logical CZ gates over distance-3 logical qubits in~\cite{ataides2025neural}, and 10 logical measurements in an LDPC code with 108 logical qubits~\cite{bhardwaj2026high}.
Streaming decoding of logical operations has been demonstrated for a magic state factory with 15 logical qubits and 27 logical measurements~\cite{bombin2023modular} and for random logical Pauli measurements over 100 distance-5 surface codes using a network of FPGA~\cite{liyanage2025network}.

In this work, we demonstrate a decoding system, running on a single CPU, capable of real-time decoding for workloads compiled over a fault-tolerant trapped-ion architecture with up to 408 logical qubits and millions of logical operations, spanning 68 LDPC memory blocks and 20 magic state factories formed using 11,680 physical qubits.
Our decoding system is based on a dual sliding-window decoder built around the beam search decoder~\cite{ye2026beam}. 
We generate DEMs on the fly and introduce memory optimizations that allow many decoder instances to execute concurrently on a single CPU.
We benchmark our decoder with a compiled quantum Hamiltonian simulation of a Heisenberg model circuit and a measurement-induced phase transition circuit~\cite{skinner2019measurement}.
We assume a SEC time varying between 1 ms and 5 ms and a noise rate between $5\times 10^{-4}$ and $10^{-4}$ which has been achievable over small trapped ion devices~\cite{loschnauer2025scalable, hughes2025trapped}.
The stretch of the computation induced by all decoding delays remains under $0.3\%$ of the total duration of the computation at noise rate $p_{\mathrm{CNOT}} = 10^{-4}$ and gracefully increases to under $12\%$ at $p_{\mathrm{CNOT}} = 5\times 10^{-4}$.

A key ingredient in the realization of a single-CPU real-time MegaQuOp decoder is the simplicity of the walking cat architecture (WCA)~\cite{tripier2026fault}.
Logical computation does not require merging or deforming qLDPC memory blocks or changing their syndrome-extraction circuits. Instead, cat-based measurements are inserted into an otherwise regular stream of syndrome extraction, while a large class of Clifford operations is handled through software frame tracking, making the entire classical decoding pipeline, from online DEM generation to decoding, feasible at MegaQuOp scale using only 12 cores of a single commodity CPU.

\begin{figure*}[t]
\centering
\begin{minipage}[t]{0.36\linewidth}
\raggedright
\makebox[\linewidth]{\panel{fig:main-a}}\\[2pt]
\begin{tikzpicture}[
  font=\small,
  >={Stealth[length=2.2mm]},
  gateH/.style={draw=black, fill=white, rounded corners=1pt,
                minimum size=3.8mm, inner sep=0pt, font=\scriptsize},
  gateT/.style={draw=black, fill=white, rounded corners=1pt,
                minimum size=3.8mm, inner sep=0pt, font=\scriptsize},
  mem/.style={draw=green!45!black, fill=green!25, rounded corners=2pt,
              minimum width=4.6mm, minimum height=4.6mm, inner sep=0pt,
              font=\scriptsize\bfseries, text=green!25!black},
  fac/.style={draw=magenta!60!black, fill=magenta!22, rounded corners=2pt,
              minimum width=4.6mm, minimum height=4.6mm, inner sep=0pt,
              font=\scriptsize\bfseries, text=magenta!40!black},
  lbl/.style={font=\scriptsize}]
\foreach \i in {0,...,5} \draw[black!70] (0,2.20-\i*0.34) -- (1.9,2.20-\i*0.34);
\foreach \i in {0,...,5} \draw[black!70] (0,-0.48-\i*0.34) -- (1.9,-0.48-\i*0.34);
\node[gray] at (0.95,0.01) {$\vdots$};
\foreach \x/\yc/\yt in {0.95/1.86/1.52, 1.50/2.20/1.86,
                        0.95/-1.16/-0.82, 1.50/-1.50/-1.84}{
  \draw (\x,\yc) -- (\x,\yt);
  \fill (\x,\yc) circle (1.5pt);
  \draw[fill=white] (\x,\yt) circle (1.6mm);
  \draw (\x-0.16,\yt) -- (\x+0.16,\yt) (\x,\yt-0.16) -- (\x,\yt+0.16);
}
\foreach \x/\ya/\yb in {0.40/1.18/0.84, 0.40/-0.48/-0.82}{
  \draw (\x,\ya) -- (\x,\yb);
  \fill (\x,\ya) circle (1.5pt);
  \fill (\x,\yb) circle (1.5pt);
}
\foreach \x/\y in {0.40/2.20, 0.95/0.50,
                   0.40/-1.50, 0.95/-2.18}
  \node[gateH] at (\x,\y) {H};
\foreach \x/\y in {1.50/1.18, 1.50/-0.48}
  \node[gateH] at (\x,\y) {S};
\node[gateT] (Tc) at (1.50,0.50) {T};
\node[gateT] (Td) at (1.50,-1.16) {T};
\draw[decorate,decoration={brace,amplitude=4pt},green!45!black] (2.15,2.38) -- (2.15,0.32);
\node[lbl,green!35!black,anchor=west,align=left] (glab1) at (2.35,1.40) {6 logical\\ qubits};
\draw[decorate,decoration={brace,amplitude=4pt},green!45!black] (2.15,-0.30) -- (2.15,-2.36);
\node[lbl,green!35!black,anchor=west,align=left] (glab2) at (2.35,-1.33) {6 logical\\ qubits};
\begin{scope}[shift={(4.35,0)}]
  \draw[rounded corners=9pt, fill=teal!6, draw=black!60, thick]
      (-0.30,-2.36) rectangle (1.98,2.40);
  \foreach \r in {0,...,4} \foreach \c in {0,...,2}
    \node[mem] (m\the\numexpr3*\r+\c\relax) at (0.23+\c*0.61, 2.05-\r*0.58) {M};
  \foreach \c in {0,1}
    \node[mem] (m\the\numexpr15+\c\relax) at (0.535+\c*0.61, -0.85) {M};
  \foreach \c in {0,...,2} \node[fac] (f\c) at (0.23+\c*0.61, -1.43) {T};
  \foreach \c in {0,1}
    \node[fac] (f\the\numexpr3+\c\relax) at (0.535+\c*0.61, -2.01) {T};
\end{scope}
\draw[->, green!45!black, thick]
  ([yshift=-1.2mm]glab1.north) to[out=45,in=185] (m0.west);
\draw[->, green!45!black, thick]
  ([yshift=-1.2mm]glab2.north) to[out=55,in=195] (m12.west);
\draw[->, magenta!60!black, semithick]
  (Tc.south east) to[bend left=10] (f0.west);
\draw[->, magenta!60!black, semithick]
  (Td.south east) .. controls (2.45,-1.48) and (3.00,-2.20) .. (f3.west);
\node[lbl,anchor=east] (fleg) at (6.33,-2.70) {CH2 magic factory};
\node[fac,minimum width=3.6mm,minimum height=3.2mm]
  (ftile) at ([xshift=-3.4mm]fleg.west) {};
\node[lbl,anchor=east] (mleg) at ([xshift=-4mm]ftile.west) {Q70 memory};
\node[mem,minimum width=3.6mm,minimum height=3.2mm]
  at ([xshift=-3.4mm]mleg.west) {};
\end{tikzpicture}
\end{minipage}\hfill%
\begin{minipage}[t]{0.60\linewidth}
\raggedleft
\makebox[\linewidth]{\panel{fig:main-b}}\\[2pt]
\begin{tikzpicture}[
    font=\small,
    sec/.style={draw=black!60, minimum width=0.50cm, minimum height=0.50cm,
                inner sep=0pt},
    note/.style={font=\scriptsize, align=center},
    dec/.style={fill=black!30, draw=black!65, rounded corners=1.5pt},
    >={Stealth[length=1.8mm]},
]
\foreach \i in {0,...,10} \node[sec,fill=blue!16] at (\i*0.50,0) {};
\foreach \i in {11,...,16}
  \node[sec,fill=red!10,draw=red!70!black,
        path picture={\pattern[pattern=north east lines,pattern color=red!60]
        (path picture bounding box.south west) rectangle
        (path picture bounding box.north east);}] at (\i*0.50,0) {};
\draw[decorate, decoration={brace, mirror, amplitude=3pt}, red!60!black]
    (5.25,-0.40) -- (8.25,-0.40)
    node[note, red!60!black, midway, below=4pt, align=center]
    {stretch episode: 6 stall SECs,\\ logical measurements frozen};
\foreach \w in {0,...,3}{
  \pgfmathsetmacro{\xs}{3*\w*0.50-0.25}
  \pgfmathsetmacro{\xe}{(3*\w+4)*0.50+0.25}
  \pgfmathsetmacro{\yb}{0.55+(3-\w)*0.65}
  \pgfmathsetmacro{\yt}{\yb+0.40}
  \fill[blue!32] (\xs,\yb) rectangle (\xe,\yt);
  \draw[blue!45!black,semithick,rounded corners=1.5pt] (\xs,\yb) rectangle (\xe,\yt);
}
\fill[dec] (2.25,2.50) rectangle (3.40,2.90);
\node[note,black!70,anchor=west] at (3.45,2.70) {no backlog};
\fill[dec] (3.75,1.85) rectangle (6.50,2.25);
\node[note,red!70!black,anchor=west] at (6.55,2.05) {2.5\,SEC backlog};
\draw[<->,red!70!black,thick,densely dashed] (5.25,1.40) -- (6.50,1.40);
\fill[dec] (6.50,1.20) rectangle (7.35,1.60);
\node[note,red!70!black,anchor=west] at (7.40,1.40) {1.2\,SEC backlog};
\draw[<->,red!70!black,thick,densely dashed] (6.75,0.75) -- (7.35,0.75);
\fill[dec] (7.35,0.55) rectangle (7.92,0.95);
\node[note,black!70,anchor=west] at (7.97,0.75) {backlog drained};
\node[sec,fill=blue!16,minimum size=0.3cm] at (0.0,-1.60) {};
\node[note,anchor=west] at (0.2,-1.60) {ordinary SEC};
\node[sec,fill=red!10,draw=red!70!black,minimum size=0.3cm,
      path picture={\pattern[pattern=north east lines,pattern color=red!60]
      (path picture bounding box.south west) rectangle
      (path picture bounding box.north east);}] at (2.6,-1.60) {};
\node[note,anchor=west] at (2.8,-1.60) {stall SEC};
\node[sec,fill=black!30,minimum size=0.3cm] at (4.6,-1.60) {};
\node[note,anchor=west] at (4.8,-1.60) {decode time};
\node[sec,fill=blue!32,minimum size=0.3cm] at (0.0,-2.10) {};
\node[note,anchor=west] at (0.2,-2.10) {decoding window (5 SECs)};
\draw[<->,red!70!black,thick,densely dashed] (4.6,-2.10) -- (5.3,-2.10);
\node[note,anchor=west] at (5.4,-2.10) {delayed start};
\end{tikzpicture}
\end{minipage}\\[12pt]
\begin{minipage}[t]{0.56\linewidth}
\raggedright
\makebox[\linewidth]{\panel{fig:main-c}}\\[2pt]
\begin{tikzpicture}[
    font=\small,
    sec/.style={draw=black!60, minimum width=0.50cm, minimum height=0.50cm,
                inner sep=0pt},
    note/.style={font=\scriptsize, align=center},
    >={Stealth[length=2.2mm]},
]
\begin{scope}[on background layer]
  \fill[blue!5]    (-0.25,-2.90) rectangle (3.25,3.45);
  \fill[orange!12] ( 3.25,-2.90) rectangle (5.75,3.45);
  \fill[blue!5]    ( 5.75,-2.90) rectangle (9.75,3.45);
\end{scope}
\node[note,font=\scriptsize\bfseries] at (4.50,3.66) {Viterbi cat region};
\foreach \i in {0,...,6}   \node[sec,fill=blue!16]   (S\i) at (\i*0.50,0) {};
\foreach \i in {7,...,11}  \node[sec,fill=orange!30] (S\i) at (\i*0.50,0) {};
\foreach \i in {12,...,19} \node[sec,fill=blue!16]   (S\i) at (\i*0.50,0) {};
\node[note,anchor=west] (sslab) at (-0.20,0.47) {syndrome stream};
\draw[->,black!70] ([xshift=1.5mm]sslab.east) -- (4.60,0.47);
\foreach \w in {0,...,5}{
  \pgfmathsetmacro{\xs}{3*\w*0.50-0.25}
  \pgfmathsetmacro{\xc}{(3*\w+2)*0.50+0.25}
  \pgfmathsetmacro{\xe}{(3*\w+4)*0.50+0.25}
  \pgfmathsetmacro{\yb}{0.40+(5-\w)*0.48}   
  \pgfmathsetmacro{\yt}{\yb+0.40}
  \fill[blue!32] (\xs,\yb) rectangle (\xc,\yt);
  \pattern[pattern=north east lines,pattern color=blue!55] (\xc,\yb) rectangle (\xe,\yt);
  \draw[blue!45!black,semithick,rounded corners=1.5pt] (\xs,\yb) rectangle (\xe,\yt);
}
\node[note,blue!40!black] at (1.50,1.50)
  {\textbf{Error decoder}\\ window 5, commit 3\\ runs continuously};
\foreach \k in {0,...,2}{
  \pgfmathsetmacro{\xs}{(7+\k)*0.50-0.25}
  \pgfmathsetmacro{\xc}{(7+\k)*0.50+0.25}
  \pgfmathsetmacro{\xe}{(8+\k)*0.50+0.25}
  \pgfmathsetmacro{\yt}{-1.18-\k*0.48}
  \pgfmathsetmacro{\yb}{\yt-0.40}
  \fill[blue!32] (\xs,\yb) rectangle (\xc,\yt);
  \pattern[pattern=north east lines,pattern color=blue!55] (\xc,\yb) rectangle (\xe,\yt);
  \draw[blue!45!black,semithick,rounded corners=1.5pt] (\xs,\yb) rectangle (\xe,\yt);
}
\node[note,blue!40!black] at (1.50,-1.60)
  {\textbf{Outcome decoder}\\ window 2, commit 1\\ runs during\\ measurements};
\draw[dashed,thick,red!55!black] (3.25,-0.25) -- (3.25,-2.56);
\node[note,red!55!black,anchor=north] at (3.25,-2.58) {spawn};
\draw[dashed,thick,red!55!black] (5.75,-0.25) -- (5.75,-2.56);
\node[note,red!55!black,anchor=north] (retlab) at (5.75,-2.58) {retire};
\foreach \i in {1,...,5}{
  \pgfmathsetmacro{\cx}{(6+\i)*0.50}
  \node[text=orange!60!black,inner sep=1pt,
        font=\scriptsize] (c\i) at (\cx,-0.90) {$c_{\i}$};
  \draw[->,orange!60!black] (\cx,-0.27) -- (c\i.north);
}
\foreach \k in {0,...,2}{
  \pgfmathsetmacro{\dx}{(8+\k)*0.50+0.62}
  \pgfmathsetmacro{\dy}{-1.38-\k*0.48}
  \node[text=blue!45!black,inner sep=1pt,
        font=\scriptsize] (d\k) at (\dx,\dy) {$d_{\the\numexpr\k+1\relax}$};
  \pgfmathsetmacro{\hx}{(7+\k)*0.50+0.25}
  \draw[->,blue!45!black] (\hx,\dy) -- (d\k.west);
}
\node[draw,thick,rounded corners=3pt,fill=blue!10,align=center,
      font=\scriptsize] (vote) at (7.80,-1.70)
      {Viterbi margin is\\
       reached if $c_1{\oplus}d_1$\\
       $=\,c_2{\oplus}d_2 =\, c_3{\oplus}d_3$};
\draw[->,blue!45!black] (d0.east) -- (vote.west);
\draw[->,blue!45!black] (d1.east) -- ($(vote.north west)!0.75!(vote.south west)$);
\draw[->,blue!45!black] (d2.east) -- (vote.south west);
\draw[->,orange!60!black] (c3.south) -- ($(vote.north west)!0.25!(vote.south west)$);
\draw[->,densely dashed,thick,red!55!black]
  (vote.south) .. controls (7.45,-2.75) .. (retlab.east)
  node[note,red!55!black,pos=0.30,right=2pt] {stop signal};
\node[sec,fill=blue!16,minimum size=0.3cm]   at (0.0,-3.16) {};
\node[note,anchor=west] at (0.2,-3.16) {ordinary SEC};
\node[sec,fill=orange!30,minimum size=0.3cm] at (2.7,-3.16) {};
\node[note,anchor=west] at (2.9,-3.16) {cat SEC};
\node[sec,fill=blue!32,minimum size=0.3cm]   at (5.0,-3.16) {};
\node[note,anchor=west] at (5.2,-3.16) {commit};
\node[sec,pattern=north east lines,pattern color=blue!55,minimum size=0.3cm] at (7.3,-3.16) {};
\node[note,anchor=west] at (7.5,-3.16) {look-ahead};
\end{tikzpicture}
\end{minipage}\hfill%
\begin{minipage}[t]{0.43\linewidth}
\centering
\panel{fig:main-d}\\[2pt]
\begin{tikzpicture}[
  cmipt/.style={color=blue!60!black,  mark=*},
  cn64/.style={color=red!70!black,    mark=square*},
  cn266/.style={color=orange!85!black,mark=diamond*},
]
\begin{axis}[
  width=7.4cm, height=6.72cm, ymode=log,
  xmin=0.7, xmax=5.3, xtick={1,2,3,4,5},
  xlabel={$p_\mathrm{CNOT}$ ($\times10^{-4}$)},
  ylabel={stretch},
  grid=both,
  major grid style={black!20, line width=0.3pt},
  minor grid style={black!10, line width=0.2pt},
  axis on top,
  tick label style={font=\scriptsize}, label style={font=\scriptsize},
  ylabel style={font=\scriptsize},
  legend style={font=\scriptsize, at={(0.5,-0.24)}, anchor=north,
                draw=none, fill=none, legend columns=2,
                /tikz/every even column/.append style={column sep=4mm}},
  mark size=1.6pt, every axis plot/.append style={semithick},
]
\addplot[cmipt] coordinates {(1,0.00236) (2,0.01121) (3,0.03228) (4,0.06672) (5,0.11531)};
\addplot[cn64]  coordinates {(1,0.00182) (2,0.00869) (3,0.02928) (4,0.05794) (5,0.10248)};
\addplot[cn266] coordinates {(1,0.00021) (2,0.00048) (3,0.00116) (4,0.00386) (5,0.00721)};
\legend{MIPT, Heisenberg n64, Heisenberg n266}
\end{axis}
\end{tikzpicture}
\end{minipage}
\caption{\textbf{(a)} Distributing a compiled Clifford~$+$~$T$ circuit onto a WCA instance. Each memory block (M) is encoded by the Q70 code and holds 6 logical qubits; each $T$ gate consumes a magic state from a CH2 factory (T). Both Q70 code and CH2 factory were introduced in \cite{tripier2026fault}.
\textbf{(b)} Anatomy of a stretch episode. The decoding time budget per
window is 3 SECs. The second window runs 2.5 SECs over budget, so the
machine inserts stall SECs until the backlog drains ($2.5\to1.2\to0$).
The inserted stall SECs form one stretch episode. \textbf{(c)} Two decoders on one block's syndrome stream.
The \emph{error decoder} runs continuously and produces the Pauli-frame
corrections. A cat SEC is a cat-based measurement followed by
an ordinary SEC. The \emph{outcome decoder} is active only during logical
measurements: its shorter look-ahead gives a lower decision latency, so it
delivers each logical outcome earlier than the error decoder could. \textbf{(d)} Stretch, as defined in Eq.~\eqref{eq:def_stretch}, versus $p_\mathrm{CNOT}$ for the three benchmark circuits. MIPT and Heisenberg n64 each contain 102 logical qubits and use a decoding-time budget of 1 ms/SEC, while Heisenberg n266 contains 408 logical qubits and uses a budget of 5 ms/SEC. All decoding benchmarks were performed on a single 2024 Apple M4 Max CPU in a MacBook Pro.}
\label{fig:main}
\end{figure*}
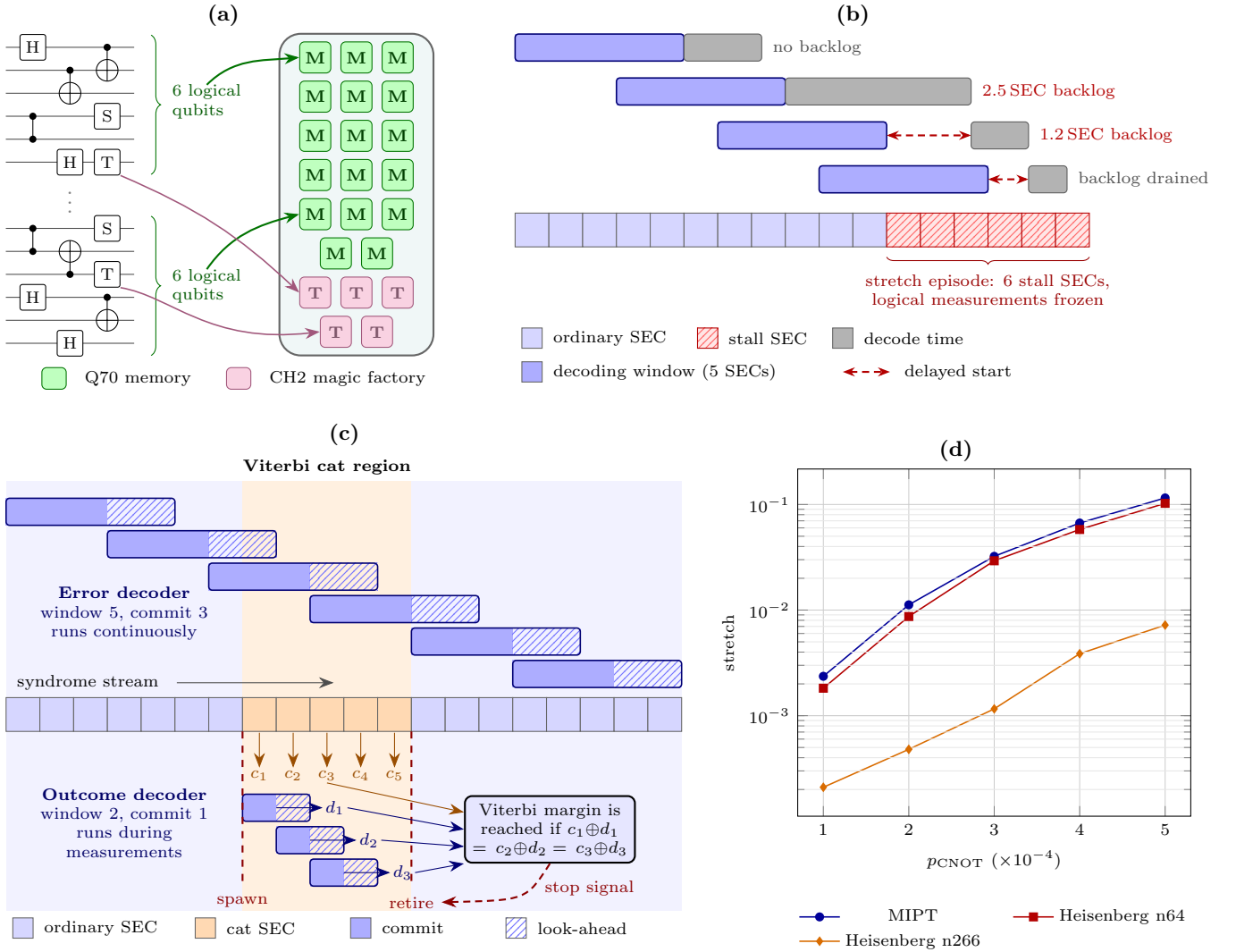

\section{Decoding the walking-cat architecture} 

\subsection{The walking-cat architecture}

We first summarize the features of the WCA~\cite{tripier2026fault} that are relevant to decoding. The WCA implements logical computation in a Pauli-based framework in which all logical operations are ultimately reduced to state preparation, frame tracking, and logical Pauli measurements~\cite{bravyi2016trading, litinski2019game}. Logical operations are divided into two classes: \emph{accessible logical Clifford gates}, which are implemented entirely in software via Clifford-frame tracking and therefore do not correspond to any physical action on the encoded data, and all remaining operations, which require physical implementation through measurement-based procedures.

In the configurations studied here, the Q70 memory block encodes six logical qubits, while the Q54 block used in the CH2 factory encodes two logical qubits. For both codes, all logical Clifford gates acting within a single block are accessible and can be implemented entirely through Clifford-frame tracking~\cite{tripier2026fault}. The main Clifford operation that cannot be handled by the frame of a single block is a CNOT between logical qubits residing in different blocks; such an inter-block CNOT is instead implemented through measurement-based procedures involving two-block logical Pauli measurements, denoted LM2~\cite{tripier2026fault}. Non-Clifford resources, in particular $T$ gates, are introduced via magic states prepared in CH2 magic-state factories and are consumed through sequences of logical Pauli measurements and measurement-conditioned updates of the Clifford frame.

Logical Pauli measurements are implemented physically using cat states. A cat state is used to measure a chosen physical representative of the logical operator and is then read out. Since an individual cat-based measurement (CM) is noisy, measurements are repeated to reach higher reliability, with a syndrome-extraction cycle separating consecutive repetitions. We call the interval consisting of a CM followed by this SEC a \emph{cat SEC}; SECs without an associated CM are \emph{ordinary SECs}.
In memory blocks of the WCA, these repeated CMs are processed using an adaptive Viterbi scheme that updates a likelihood over measurement histories and terminates once a prescribed confidence threshold is reached. In contrast, magic-state factory blocks use the error-detected measurement (EDM) scheme, which post-selects on agreement of all constituent CMs before accepting the outcome.

This organization makes the WCA particularly simple from the viewpoint of decoding. Logical computation does not deform or merge the encoded blocks, nor does it change their syndrome-extraction circuits. Instead, each live block produces a continuous stream of syndrome data consisting of ordinary SECs punctuated by known cat SECs. Thus the decoding problem retains the structure of QEC memory experiments while logical measurements appear as a sparse overlay on that otherwise regular stream. The decoder constructions below exploit precisely this structure.

\subsection{Dual-decoder structure}
We propose a dual-decoder system to decode the WCA, in which each block's syndrome stream is processed by two sliding-window decoders with complementary roles, as illustrated in Fig.~\ref{fig:main-c}. The \emph{error decoder} runs continuously over both ordinary and cat SECs. It uses a sliding window of 5 SECs with a 3-SEC commit region and produces the Pauli-frame corrections used throughout the computation. The \emph{outcome decoder}, by contrast, is active only during a logical measurement. When a logical measurement begins, the outcome decoder is launched within the committed region of the error decoder and is initialized with the Pauli frame committed by the error decoder up to that SEC. This ensures that errors accumulated before the measurement are correctly accounted for when interpreting the subsequent cat-state readouts. It uses a smaller window of 2 SECs with a 1-SEC commit region, allowing it to correct successive cat-state readouts with lower latency. 
For an inter-block LM2, the corrected cat-state readouts produced by the outcome decoders of the two participating blocks are combined to form a joint LM2 readout before applying the Viterbi procedure; the runtime implementation is described in Sec.~\ref{sec:logical_control}. The resulting corrected readouts, either from a single block or from an LM2 across two blocks, are then used to determine when a Viterbi measurement should terminate, or whether an EDM in a factory block should be accepted or rejected. Once the logical measurement terminates, the outcome decoder retires. The outcome decoder lies directly on the execution critical path because logical measurement outcomes determine the subsequent logical operations.  The computation therefore cannot proceed until the current measurement has terminated and its decoded outcome is available.  For each cat-state readout, the outcome decoder must determine whether the Viterbi stopping condition has been reached or the EDM should be accepted or rejected, and provide the corresponding logical measurement result to the controller.  If this decision is not available by the next SEC boundary, the measurement must continue for an additional cycle, delaying all subsequent operations that depend on its outcome.  Outcome-decoding latency can therefore translate directly into additional executed SECs, motivating the use of a dedicated low-latency decoder for this task.

The two decoding tasks, however, have very different accuracy requirements. For cat-state readout, there exist pairs of error mechanisms that produce exactly the same syndrome but differ in whether they flip the CM outcome. No decoder, regardless of its accuracy, can distinguish between such mechanisms from the syndrome data alone. They therefore impose an intrinsic error floor on the corrected cat-state readout. These intrinsic errors cannot be suppressed by improving the decoder, but they can be suppressed at the measurement-protocol level through Viterbi voting or EDM repetition.

For the outcome decoder, it is therefore sufficient to suppress the remaining, correctable contribution well below this intrinsic floor; in our implementation, we target an error rate of the correctable contribution below approximately one tenth of the intrinsic contribution. The error decoder, by contrast, suppresses the correctable contribution to below approximately one thousandth of the intrinsic error rate. Such accuracy is necessary for maintaining the Pauli frame over the full computation, but is unnecessary for determining individual logical-measurement outcomes and comes at the cost of substantially higher latency. The shorter-window outcome decoder therefore reduces the latency on the measurement critical path while introducing only a negligible increase in the total readout error, which remains dominated by the intrinsic contribution.

\subsection{On-the-fly DEM generation with a static Tanner graph}

A sliding-window decoder operates on a detector error model, which specifies how elementary error mechanisms affect the observed detectors. A DEM consists of two components: a Tanner graph, equivalently a parity-check matrix, that records which detectors are triggered by each error mechanism, and a vector of prior probabilities for those mechanisms~\cite{gidney2021stim}. In a memory experiment, all noise arises from repeated SECs, so the DEM is identical from one decoding window to the next, apart from the first and last windows where temporal boundary conditions differ.

Compared with a memory experiment, WCA adds CMs to an otherwise unchanged sequence of SECs.  As described above, each CM is followed by an ordinary SEC, and we treat this pair as a single \emph{cat SEC}.  A decoding window containing a cat SEC therefore has a different DEM from a memory window. Moreover, the error mechanisms introduced by a CM depend on the logical Pauli operator being measured, since different operators couple the cat state to different sets of data qubits. Thus, different cat SECs generally correspond to different DEMs.  Naively, accommodating these changes would require modifying the Tanner graph, or parity-check matrix, for each pattern of cat SECs encountered during execution. Such real-time changes to the decoding graph would be prohibitively expensive, as they require restructuring the message-passing topology and associated decoder data structures during execution.

The structure of the WCA allows us to avoid this reconstruction entirely.  We observe that, for every error mechanism introduced by a CM, there is an error mechanism in the accompanying SEC that triggers exactly the same set of detectors.  The two mechanisms therefore correspond to identical columns of the parity-check matrix and can be represented by a single error variable with a modified prior.  If two independent error mechanisms with identical detector signatures occur with probabilities $p_1$ and $p_2$, their combined effect is an error whenever exactly one of them occurs, giving the combined prior $p_{\mathrm{comb}} = p_1(1-p_2)+p_2(1-p_1)$.
Thus, incorporating a CM requires changing only the priors of the affected error variables; the parity-check matrix itself is unchanged.

As a result, the same Tanner graph can be reused for all bulk decoding windows, whether they contain cat SECs or not. This allows us to generate the window-specific DEM on the fly: when a cat SEC enters the decoding window, we update only the corresponding entries of the prior vector, and restore the ordinary-memory priors when it leaves. No reconstruction of the Tanner graph or message-passing topology is required. The resulting DEM generation is sufficiently lightweight to be performed in software alongside decoding on the same CPU. Thus, the entire WCA syndrome stream can be processed using a fixed Tanner graph with dynamically updated priors. This simplification is a direct consequence of the WCA architecture: logical computation adds CMs to otherwise unchanged syndrome extraction, without requiring operations such as lattice surgery or code deformation.

\subsection{Decoding time budgets and stretch}

Let $\tau_{\mathrm{SEC}}$ denote the physical duration of one SEC. A sliding-window decoder with a commit size of $c$ SECs receives a new window every $c\tau_{\mathrm{SEC}}$ and therefore has a nominal decoding budget of $c\tau_{\mathrm{SEC}}$ per window.  The error decoder, with a 3-SEC commit region, has a budget of $3\tau_{\mathrm{SEC}}$ per window, whereas the outcome decoder, with a 1-SEC commit region, has a budget of $\tau_{\mathrm{SEC}}$.  For example, at $\tau_{\mathrm{SEC}}=1$ ms these budgets are 3 ms and 1 ms, respectively.

Exceeding these budgets affects the two decoders differently.  Recall that, when a logical measurement begins, the outcome decoder must be initialized with the Pauli frame produced by the error decoder up to that SEC. The error decoder must therefore remain caught up with the syndrome stream so that this frame is available when the measurement begins.  If decoding a window takes longer than its budget, the error decoder accumulates a \emph{decoding backlog}, equal to the amount by which its decoding work has fallen behind the incoming syndrome stream.  Subsequent windows that finish within budget can reduce this backlog. If a code block still has decoding backlog when it is scheduled to perform a cat-based measurement, that measurement is postponed until the error decoder catches up and the backlog is drained. During this delay, ordinary syndrome extraction continues, inserting additional SECs before cat-based measurements can start (Fig.~\ref{fig:main-b}). Because all code blocks advance on a synchronized SEC clock, this delay extends the global schedule: additional SECs are inserted for every block, with logical measurements paused and ordinary syndrome extraction continuing until the backlog is drained.

The outcome decoder extends the schedule through a different mechanism.  Its decoded readouts determine when a Viterbi measurement terminates and whether an EDM is accepted or rejected.  If the required decoding result is not available by the next SEC boundary, the corresponding decision is delayed. For a Viterbi measurement, the absence of a stopping decision causes additional cat SECs to be executed.  For an EDM, ordinary SECs continue while the decoder catches up and produces the accept/reject decision. Thus, both error-decoder backlog and outcome-decoder latency  increase the number of SECs executed in the computation.

We quantify this increase by the \emph{stretch}.  Let $N_{\mathrm{baseline}}$ denote the number of SECs that the computation would execute with zero decoder latency, and let $N_{\mathrm{stretch}}$ denote the additional SECs executed because decoding does not complete within its real-time budget. The total number of executed SECs is then $N_{\mathrm{executed}} = N_{\mathrm{baseline}} + N_{\mathrm{stretch}}$, and we define
\begin{equation} \label{eq:def_stretch}
    \mathrm{stretch}
    = \frac{N_{\mathrm{stretch}}}{N_{\mathrm{baseline}}}.
\end{equation}
Here, $N_{\mathrm{stretch}}$ includes both SECs inserted to clear error-decoder backlog and SECs added by delayed outcome-decoder decisions. We refer to a contiguous sequence of such additional SECs as a \emph{stretch episode}. Since stretch episodes can vary widely in duration, we report both the total stretch, shown in Fig.~\ref{fig:main-d}, and the stretch-episode length statistics in Tables~\ref{tab:mipt}--\ref{tab:n266}.

\section{Real-time decoding of MegaQuOp workloads}

\begin{table*}[t]
\centering
{\footnotesize\setlength{\tabcolsep}{4pt}%
\begin{tabular}{l|c|r|r|r|r|r}
  circuit & WCA instance & \shortstack{logical\\ qubits} &
  $T$ gates & \shortstack{logical\\ measurements} &
  \shortstack{total SECs\\ (all blocks)} & \shortstack{decoding\\ time budget} \\\hline
  MIPT            & $17{\times}\mathrm{Q70}+5{\times}\mathrm{CH2}$  & 102 & 1{,}087{,}434 & 1{,}100{,}227 & 24{,}755{,}302 & 1 ms/SEC \\
  Heisenberg n64  & $17{\times}\mathrm{Q70}+5{\times}\mathrm{CH2}$  &  102 &   139{,}408   &   327{,}066   &  6{,}965{,}200 & 1 ms/SEC \\
  Heisenberg n266 & $68{\times}\mathrm{Q70}+20{\times}\mathrm{CH2}$ & 408 &   555{,}130   & 1{,}318{,}310 & 31{,}548{,}792 & 5 ms/SEC \\
\end{tabular}}
\caption{Benchmark circuit inventory}
\label{tab:inventory}
\end{table*}

\begin{table*}[t]
\centering
\footnotesize
\begin{tabular}{c|c|c|c|c|c|c|c|c|c}
$p_\mathrm{CNOT}$ & stretch & \multicolumn{2}{c|}{\shortstack{stretch\\ episode length}} & \multicolumn{3}{c|}{\shortstack{error decoder\\ time per window (ms)}} & \multicolumn{3}{c}{\shortstack{outcome decoder\\ time per SEC (ms)}} \\\cline{3-10}
($\times10^{-4}$) &  & \makebox[10mm]{mean} & \makebox[10mm]{p99.9} & \makebox[10mm]{mean} & \makebox[10mm]{p99.9} & \makebox[12mm]{$>3$\,ms} & \makebox[10mm]{mean} & \makebox[10mm]{p99.9} & \makebox[12mm]{$>1$\,ms} \\\hline
1 & 0.24\% & 2.8 & 31 & 0.99 & 3.32 & 0.26\% & 0.31 & 1.35 & 0.18\% \\
2 & 1.12\% & 3.1 & 41 & 1.19 & 5.25 & 1.26\% & 0.33 & 2.38 & 0.69\% \\
3 & 3.23\% & 3.2 & 42 & 1.40 & 6.85 & 3.76\% & 0.37 & 2.83 & 1.52\% \\
4 & 6.67\% & 3.8 & 64 & 1.63 & 9.47 & 6.58\% & 0.41 & 3.40 & 2.72\% \\
5 & 11.53\% & 4.3 & 105 & 1.89 & 13.68 & 10.70\% & 0.45 & 4.05 & 3.92\% \\
\end{tabular}
\caption{\textbf{MIPT circuit (102 logical qubits, 22 blocks; budget 1\,ms/SEC).}\ Decoding is performed on a single 2024 Apple M4 Max CPU using 12 cores, with 4 cores allocated to the outcome decoder and 8 to the error decoder. All decoder timing statistics include both on-the-fly DEM generation and beam search decoding. $N_{\mathrm{baseline}}$ (total number of SECs over all 22 blocks in the zero-decoder-latency schedule) ranges from 24,755,302 to 27,302,022 SECs across the sweep of $p_\mathrm{CNOT}$.}
\label{tab:mipt}
\end{table*}

\begin{table*}[t]
\centering
\footnotesize
\begin{tabular}{c|c|c|c|c|c|c|c|c|c}
$p_\mathrm{CNOT}$ & stretch & \multicolumn{2}{c|}{\shortstack{stretch\\ episode length}} & \multicolumn{3}{c|}{\shortstack{error decoder\\ time per window (ms)}} & \multicolumn{3}{c}{\shortstack{outcome decoder\\ time per SEC (ms)}} \\\cline{3-10}
($\times10^{-4}$) &  & \makebox[10mm]{mean} & \makebox[10mm]{p99.9} & \makebox[10mm]{mean} & \makebox[10mm]{p99.9} & \makebox[12mm]{$>3$\,ms} & \makebox[10mm]{mean} & \makebox[10mm]{p99.9} & \makebox[12mm]{$>1$\,ms} \\\hline
1 & 0.18\% & 2.7 & 24 & 1.00 & 3.38 & 0.31\% & 0.26 & 0.72 & 0.06\% \\
2 & 0.87\% & 3.3 & 67 & 1.18 & 5.19 & 1.20\% & 0.28 & 1.74 & 0.27\% \\
3 & 2.93\% & 3.2 & 44 & 1.39 & 6.84 & 3.75\% & 0.31 & 2.75 & 1.05\% \\
4 & 5.79\% & 3.8 & 67 & 1.63 & 9.28 & 6.67\% & 0.33 & 2.73 & 1.41\% \\
5 & 10.25\% & 4.1 & 85 & 1.85 & 11.62 & 10.48\% & 0.37 & 3.65 & 2.54\% \\
\end{tabular}
\caption{\textbf{Heisenberg n64 (102 logical qubits, 22 blocks; budget 1\,ms/SEC).}\ Decoding is performed on a single 2024 Apple M4 Max CPU using 12 cores, with 4 cores allocated to the outcome decoder and 8 to the error decoder. All decoder timing statistics include both on-the-fly DEM generation and beam search decoding. $N_{\mathrm{baseline}}$ (total number of SECs over all 22 blocks in the zero-decoder-latency schedule) ranges from 6,965,200 to 7,587,932 SECs across the sweep of $p_\mathrm{CNOT}$.}
\label{tab:n64}
\end{table*}

\begin{table*}[t]
\centering
\footnotesize
\begin{tabular}{c|c|c|c|c|c|c|c|c|c}
$p_\mathrm{CNOT}$ & stretch & \multicolumn{2}{c|}{\shortstack{stretch\\ episode length}} & \multicolumn{3}{c|}{\shortstack{error decoder\\ time per window (ms)}} & \multicolumn{3}{c}{\shortstack{outcome decoder\\ time per SEC (ms)}} \\\cline{3-10}
($\times10^{-4}$) &  & \makebox[10mm]{mean} & \makebox[10mm]{p99.9} & \makebox[10mm]{mean} & \makebox[10mm]{p99.9} & \makebox[12mm]{$>15$\,ms} & \makebox[10mm]{mean} & \makebox[10mm]{p99.9} & \makebox[12mm]{$>5$\,ms} \\\hline
1 & 0.02\% & 10.9 & 32 & 3.51 & 7.53 & 0.01\% & 0.79 & 2.36 & 0.00\% \\
2 & 0.05\% & 5.4 & 51 & 4.02 & 10.36 & 0.02\% & 0.81 & 3.44 & 0.01\% \\
3 & 0.12\% & 4.4 & 52 & 4.61 & 14.71 & 0.10\% & 0.84 & 4.17 & 0.03\% \\
4 & 0.39\% & 8.1 & 252 & 5.22 & 20.39 & 0.20\% & 0.89 & 4.76 & 0.08\% \\
5 & 0.72\% & 6.7 & 153 & 5.97 & 41.86 & 0.50\% & 0.96 & 5.34 & 0.15\% \\
\end{tabular}
\caption{\textbf{Heisenberg n266 (408 logical qubits, 88 blocks; budget 5\,ms/SEC).}\ Decoding is performed on a single 2024 Apple M4 Max CPU using 12 cores, with 4 cores allocated to the outcome decoder and 8 to the error decoder. All decoder timing statistics include both on-the-fly DEM generation and beam search decoding. $N_{\mathrm{baseline}}$ (total number of SECs over all 88 blocks in the zero-decoder-latency schedule) ranges from 31,548,792 to 38,704,160 SECs across the sweep of $p_\mathrm{CNOT}$.}
\label{tab:n266}
\end{table*}

\subsection{Benchmark circuits}

We benchmark the real-time decoder on three fault-tolerant workloads spanning two circuit families and system sizes from 102 to 408 logical qubits: a measurement-induced phase transition (MIPT) circuit which could make an interesting quantum information experiment on a large-scale quantum computer and two disordered-Heisenberg circuits, denoted Heisenberg n64 and Heisenberg n266, illustrating quantum Hamiltonian simulation.

The MIPT circuit realizes the measurement-induced phase-transition model of~\cite{skinner2019measurement}. It consists of 40 one-dimensional brickwork layers of Haar-random two-qubit gates on the logical data qubits, with each data qubit measured in the $Z$ basis with probability $p_{\mathrm{mipt}}=0.16$ after every layer. After compilation, the circuit occupies 102 logical qubits, including both data qubits and ancilla qubits introduced by the compiler. Each Haar-random two-qubit gate is compiled to Clifford$+T$ using the KAK decomposition and Ross--Selinger synthesis~\cite{ross2016optimal}, with operator-norm error at most $0.01$. The resulting decompositions require 488--580 $T$ gates per two-qubit gate, with a mean of 538. Under the WCA execution, accessible logical Clifford gates are implemented through frame tracking, while each $T$ gate consumes a magic state through a joint memory--factory CM. The circuit is mapped onto $17\times\mathrm{Q70}$ memory blocks and $5\times\mathrm{CH2}$ magic-state factories; see Fig.~\ref{fig:main-a} for an illustration.

The other two benchmarks simulate dynamics of the disordered Heisenberg Hamiltonian of Eq.~(51) in~\cite{tripier2026fault} on degree-three random regular graphs. Both implement a single sixth-order Trotter step and are compiled using the logical compiler of~\cite{tripier2026fault}. Heisenberg n64 contains 64 physical-model sites and is mapped onto $17\times\mathrm{Q70}$ memory blocks and $5\times\mathrm{CH2}$ factories, using all 102 logical qubits available in the memory blocks, including ancillae introduced by the compiler. Heisenberg n266 contains 266 sites and is mapped onto $68\times\mathrm{Q70}$ memory blocks and $20\times\mathrm{CH2}$ factories, using all 408 logical qubits in the memory blocks. The two instances therefore probe the same structured workload at substantially different machine scales.

Table~\ref{tab:inventory} summarizes the WCA configuration and logical-operation counts for all three benchmarks. Because the notion of a ``logical gate'' in the MegaQuOp criterion~\cite{preskill2025beyond} depends on the chosen accounting convention, we report both the number of $T$ gates and the total number of logical measurements. The MIPT circuit contains more than one million $T$ gates and more than one million logical measurements, while Heisenberg n266 contains more than one million logical measurements. We also report the total number of SECs executed across all code blocks in the zero-decoder-latency schedule. Since all blocks advance on a synchronized SEC clock, this SEC volume is the schedule depth multiplied by the number of blocks; at $p_{\mathrm{CNOT}}=10^{-4}$, it exceeds $10^6$ SECs for all three benchmarks.

\subsection{Runtime logical control and measurement processing}
\label{sec:logical_control}

Our end-to-end simulation also implements the classical logical-control operations required by the WCA. For each code block, the controller maintains a logical Clifford frame $U$, initialized as the identity. An accessible logical Clifford gate $V$, including a conditional Clifford triggered by a logical-measurement outcome, is implemented by updating the frame as $U\leftarrow VU$ rather than applying a physical logical gate. When a logical Pauli operator $P$ is subsequently measured, the controller computes $U^\dagger P U$ and uses a precomputed lookup table to select a low-weight physical representative, $\overline{P}=\Phi_{\mathcal B}(U^\dagger P U)$, following the construction of~\cite{tripier2026fault}. For Q70 and Q54, these tables contain $4^6-1=4095$ and $4^2-1=15$ nontrivial logical Pauli operators, respectively.

For an inter-block LM2 between blocks $A$ and $B$, let $c_i^{(A)}$ denote the parity of the cat-qubit measurement outcomes associated with the support in block $A$ during round $i$, and let $d_i^{(A)}$ denote the correction inferred by the outcome decoder for that support. Define $c_i^{(B)}$ and $d_i^{(B)}$ analogously for block $B$. The corrected LM2 readout for round $i$ is $r_i^{(AB)} = c_i^{(A)}\oplus d_i^{(A)} \oplus c_i^{(B)}\oplus d_i^{(B)}$. The corrected joint readouts $r_i^{(AB)}$ are then processed by the applicable logical-measurement protocol, such as Viterbi or EDM, to determine the logical outcome and the corresponding stopping or acceptance condition.

\subsection{Real-time decoding model and implementation}

We simulate all three benchmark workloads using the circuit-level noise model of~\cite{tripier2026fault}, in which both syndrome extraction and CM are noisy. For simplicity, we do not simulate the whole cat state factory because the decoder does not play any role in this factory, but we simulate cat-based measurements at the circuit level, including two-qubit depolarizing noise with rate $p_{\mathrm{CNOT}}$ applied to each cat-state qubit and its coupled data qubit. 

The CH2 magic state factories, by contrast, are simulated at the circuit level and decoded as part of our real-time decoding demonstration. All logical operations, syndrome extraction, and logical measurements within the factories are included. To enable large-scale stabilizer simulation, we replace the physical magic state $|H\rangle$ used for the initial logical magic state preparation by the stabilizer state $|+\rangle$. This substitution is justified because our goal is to benchmark the real-time decoding workload rather than the output fidelity of the magic state factory.

The simulation also includes the runtime logical-control operations described above, including Clifford-frame updates, logical-operator conjugation, and physical-representative lookup. These operations belong to the compiler and logical-control layer rather than the decoder and are therefore not included in the decoder timing measurements reported below.

All decoding benchmarks are performed on a single 2024 Apple M4 Max CPU in a MacBook Pro. We use 12 of its 16 CPU cores, assigning eight cores to the error decoder and four to the outcome decoder. Each decoder core is assigned a fixed subset of code blocks and processes those blocks sequentially, while different cores execute concurrently. All runtime processing required by the decoder, including on-the-fly generation of the window-specific DEMs and beam search decoding, is performed on this same CPU. The timing measurements reported below include both DEM generation and decoding, as well as the cache and memory-bandwidth contention arising from running all 12 decoder processes concurrently.

We consider two hardware SEC durations. For MIPT and Heisenberg n64, each of which uses 22 code blocks and 102 logical qubits, we take $\tau_{\mathrm{SEC}}=1$ ms. The corresponding decoding budgets are 3 ms per error-decoder window and 1 ms per outcome-decoder window. For Heisenberg n266, which uses 88 code blocks and 408 logical qubits, we take $\tau_{\mathrm{SEC}}=5$ ms, giving budgets of 15 ms and 5 ms for the error and outcome decoders, respectively. The larger SEC duration compensates for the larger number of code blocks assigned to each core while keeping the same 12-core decoding allocation. We use 1 ms/SEC as a representative long-term trapped-ion hardware timescale and 5 ms/SEC as a representative near-term timescale.

Within each sliding window, both the error and outcome decoders use the beam search decoder of~\cite{ye2026beam}. The outcome decoder uses the \texttt{beam8\_230iters} configuration, whereas the error decoder uses the more accurate \texttt{beam32\_340iters} configuration, following Table~I of~\cite{ye2026beam}. To make these decoders efficient when many instances run concurrently, we reformulate how belief-propagation state is stored between beam search rounds. In the original implementation, every retained candidate stores a complete copy of the Tanner-graph edge messages at the end of each round, which are then used to initialize masked belief propagation in the next round. This state is expensive because the number of Tanner-graph edges substantially exceeds the number of error nodes.

Instead, for each retained candidate we store only the posterior log-likelihood ratio (LLR) of each error node. Let an error node have posterior LLR $\lambda$, prior LLR $\lambda_0$, and degree $d$. Since $\lambda-\lambda_0$ is the sum of its incoming check-to-variable messages, we approximate each incoming message by their average, $(\lambda-\lambda_0)/d$, and reinitialize each outgoing variable-to-check message as $\lambda-\frac{\lambda-\lambda_0}{d}$. This reduces the stored state from one value per Tanner-graph edge to one value per error node, lowering the memory required per beam candidate by approximately the average node degree. We verified that this reformulation produces similar logical error rate as the original decoder. Relative to the original beam search implementation~\cite{beam_search_github}, we further apply several implementation-level memory optimizations. Together, these changes reduce the decoder memory footprint by more than an order of magnitude and substantially reduce memory traffic, thereby mitigating memory-bandwidth contention when all 12 decoder processes run concurrently on a single CPU.

\subsection{Results}

Figure~\ref{fig:main-d} shows the stretch as a function of $p_{\mathrm{CNOT}}$ for all three benchmark workloads, with detailed decoding-time and stretch-episode statistics reported in Tables~\ref{tab:mipt}--\ref{tab:n266}. At $p_{\mathrm{CNOT}}=10^{-4}$, the stretch remains below $0.3\%$ for all three workloads, demonstrating that a single CPU can perform both on-the-fly DEM generation and decoding while keeping pace with the fault-tolerant execution.

As $p_{\mathrm{CNOT}}$ increases, the decoding-time distributions develop heavier tails and budget overruns become more frequent, leading to larger stretch. Nevertheless, the degradation is gradual: even at $p_{\mathrm{CNOT}}=5\times10^{-4}$, the stretch remains below $12\%$ for all three workloads.

Rare convergence failures can also occur for sufficiently difficult decoding windows, particularly for larger circuits or higher physical error rates. We treat these separately from stretch: a budget overrun delays the schedule but eventually produces a decoding result, whereas a convergence failure requires restarting the affected computation. Such convergence failures remain rare as long as the total SEC volume of the computation is well below the inverse of the logical error rate per SEC.

Taken together, these results show that real-time decoding is not a bottleneck at MegaQuOp scale: a single CPU is sufficient for the workloads studied here, spanning 102--408 logical qubits at realistic physical error rates. They also suggest a simple provisioning rule: decoding compute resources scale primarily with the number of code blocks assigned to each core, providing a natural path to larger machines by increasing the number of conventional CPU cores or CPUs.

\section*{Acknowledgment}

The authors thank Felix Tripier, Nolan Coble, Jacob Young, Finn Buessen, John Gamble and the whole IonQ team for their help with simulations and insightful discussions.

\bibliography{references}

\end{document}